\documentclass[11pt]{article}

\usepackage[preprint]{acl}

\usepackage{times}
\usepackage{latexsym}

\usepackage[T1]{fontenc}
\usepackage[utf8]{inputenc}

\usepackage{microtype}

\usepackage{inconsolata}

\usepackage{graphicx}
\usepackage{subcaption}

\usepackage{booktabs}
\usepackage{tabularx}
\usepackage{array}
\usepackage{enumitem}
\usepackage{amsmath}
\newcolumntype{Y}{>{\raggedright\arraybackslash}X}

\usepackage{multirow}

\title{Experience-Sensitive Game Learning:\\A Behavioral Study of Humans and Language Agents}

\author{
Yingying Guo$^{1}$\thanks{Equal contribution.} \quad
Zhuoxuan Ju$^{2}$\footnotemark[1] \quad
Ruibo Ming$^{3}$ \quad
Ruicheng Feng$^{4}$ \quad
Jinjin Gu$^{3}$ \\
$^{1}$The Chinese University of Hong Kong, Shenzhen \quad$^{2}$Georgetown University \\
$^{3}$INSAIT, Sofia University ``St. Kliment Ohridski'' \quad $^{4}$Tencent
}

\begin{document}
\maketitle
\begin{abstract}
Large language model agents are increasingly evaluated through games, but most benchmarks emphasize final outcomes rather than how players learn from repeated interaction.
We study \textit{experience-sensitive game learning}: how gameplay experience changes the decision-making behavior of humans and language agents.
We formulate experience-sensitive game learning as a framework for analyzing behavioral change across repeated gameplay, rather than only final score or win rate.
We introduce a suite of interactive games with reusable strategic structure, together with cross-game greedy-to-global metrics and game-specific behavioral diagnostics that make experience-driven change observable from action traces.
We also collect repeated-game trajectories from human players and evaluate recent self-evolving language agents in the same behavioral metric space.
Our results show that human players exhibit interpretable and relatively stable shifts from locally greedy heuristics toward more global strategic decisions.
In contrast, current self-evolving agents often show noisy and transient gains, suggesting that existing self-evolution methods remain limited in converting gameplay experience into durable changes in decision-making behavior.
\end{abstract}

\section{Introduction}
Games have long served as testbeds for intelligence.
For large language model (LLM) agents, games are especially appealing because they combine explicit rules, sequential decision-making, repeatable episodes, and rich interaction traces.
Success in such games can arise from different sources of competence.
One source is strong within-episode computation: the ability to reason from the rules, evaluate states, and search over possible actions in a particular game trajectory.
Another source is accumulated gameplay experience: repeated interaction may help a player discover reusable strategies, abandon misleading heuristics, and transfer lessons from earlier episodes to later decisions.
Although these two sources of competence are closely related in practice, they are conceptually different.
Existing studies on game-based LLM agent evaluation have largely emphasized within-episode competence, treating games as tests of reasoning, planning, or task completion within individual episodes.
Much less attention has been paid to the second form: whether repeated interaction changes the way an agent evaluates states, selects actions, and reuses experience across episodes.

In this work, we study \textit{experience-sensitive game learning}.
We use this term to describe settings in which repeated gameplay can change a player's future decisions by inducing reusable strategic knowledge.
In an experience-sensitive game, effective play is not determined only by reasoning from the rules.
It also depends on whether the player can discover patterns, revise misleading heuristics, learn and reuse lessons from prior episodes.
This formulation applies to both language agents and human players.
Our goal is to analyze how gameplay experience changes observable decision-making behavior.

Consider Othello as a motivating example.
A novice player may look at a board state and prefer the move that flips the largest number of discs in the current turn, since flipping more discs appears to make immediate progress under a simple reading of the rules.
However, with experience, players often learn that such locally greedy moves can be strategically weak, as they may reduce future mobility, expose unstable frontier discs, or give the opponent access to valuable positions.

The same board state is, therefore, interpreted differently before and after gaining experience.
What changes is not merely the player's final outcome, but the heuristic used to evaluate actions.
This makes Othello an example of experience-sensitive learning: repeated gameplay can reshape which features of a state are considered important.

To study this form of learning, we instantiate our framework with a suite of experience-sensitive games selected for their reusable strategic structure and rich behavioral traces.
For each game, we define diagnostic behavioral metrics that capture interpretable aspects of play.
These metrics are not the goal of the study but instruments for making experience-driven change observable.
They allow us to ask whether a player becomes less greedy, more exploratory, more consistent, or more capable of long-horizon planning as gameplay experience accumulates.
By turning raw gameplay trajectories into behavioral profiles, we can analyze learning dynamics within a player and compare learning patterns across different types of players.

Another component of our study is human gameplay data.
We do not use humans merely as a performance ceiling or a source of final scores.
Rather, we treat human players as learning subjects whose behavioral changes can be analyzed using the same metrics.
Human trajectories provide a reference for what experience-sensitive learning looks like in practice: which heuristics disappear, which strategies emerge, and which behavioral metrics change as players become more familiar with a game.
This allows us to compare not only whether modern agents perform as well as humans, but also whether they improve in similar ways.

Our contributions are threefold:
(1) We formulate experience-sensitive game learning as a framework for studying how humans and language agents change their behavior through repeated gameplay;
(2) We introduce a suite of experience-sensitive games and game-specific behavioral metrics that make experience-driven behavioral change observable beyond final performance;
and (3) we collect human gameplay trajectories and evaluate recent self-evolving language agents in the same metric space, enabling us to analyze human learning, agent self-improvement, and human-agent behavioral differences.
Together, our work turns games from static tests of task success into interactive environments for studying how experience reshapes decision-making behavior.

\begin{table*}[t]
  \centering
  \scriptsize
  \setlength{\tabcolsep}{5pt}
  \renewcommand{\arraystretch}{1.18}
  \begingroup
  \renewcommand{\tabularxcolumn}[1]{m{#1}}
  \begin{tabularx}{\textwidth}{@{}
      >{\raggedright\arraybackslash}m{0.12\textwidth}
      >{\centering\arraybackslash}m{0.16\textwidth}
      YYY@{}}
    \toprule
    \textbf{Game}
    & \textbf{Interface}
    & \textbf{Greedy Baseline}
    & \textbf{Global Evaluator}
    & \textbf{Learning Signal} \\
    \midrule

    \textbf{Four in a Row}
    \newline
    \emph{Adversarial search}
    &
    \includegraphics[width=0.80\linewidth]{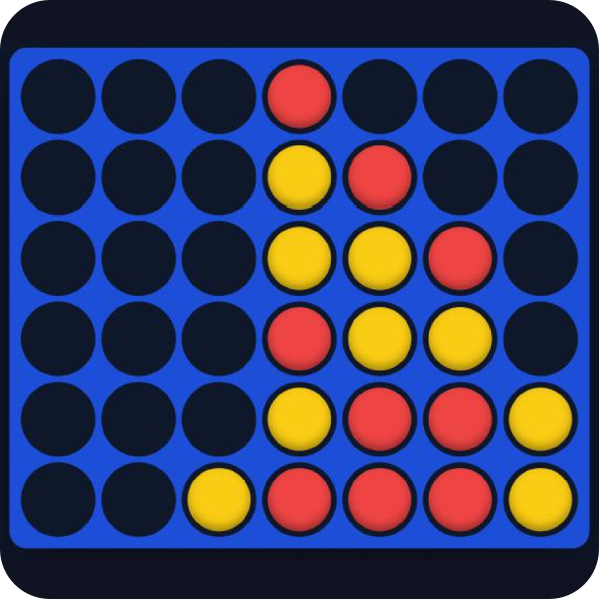}
    &
    \emph{Local line completion}: extend or block the most visible line.
    This action can look tactically useful but may miss forced threats, multi-step threats, or center control.
    &
    \textbf{Solver value} over legal columns.
    The evaluator rewards moves that are strong under global adversarial search rather than only locally visible patterns.
    &
    \textbf{Pattern completion $\rightarrow$ solver-aligned search.}
    Improvement means moving from obvious local connections toward moves that match globally strong play.
    \\

    \midrule

    \textbf{Othello6}
    \newline
    \emph{Greedy-trap strategy}
    &
    \includegraphics[width=0.80\linewidth]{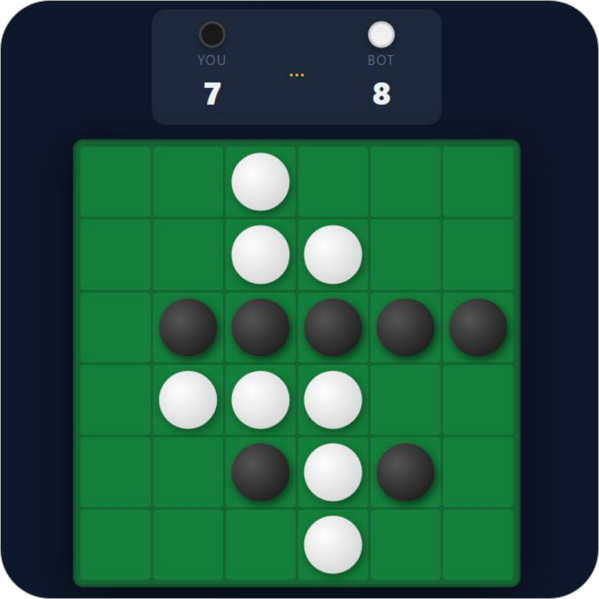}
    &
    \emph{Maximum-flip move}: choose the legal action that flips the most discs immediately.
    This reward proxy is visually salient but can reduce future mobility or expose unstable positions.
    &
    \textbf{Search value} from depth-limited minimax or related evaluation.
    The evaluator favors mobility-aware and positionally robust moves over immediate disc gain.
    &
    \textbf{Flip maximization $\rightarrow$ mobility control.}
    Improvement means avoiding short-sighted flips when they harm long-term positional value.
    \\

    \midrule

    \textbf{CircleCat}
    \newline
    \emph{Spatial containment}
    &
    \includegraphics[width=0.80\linewidth]{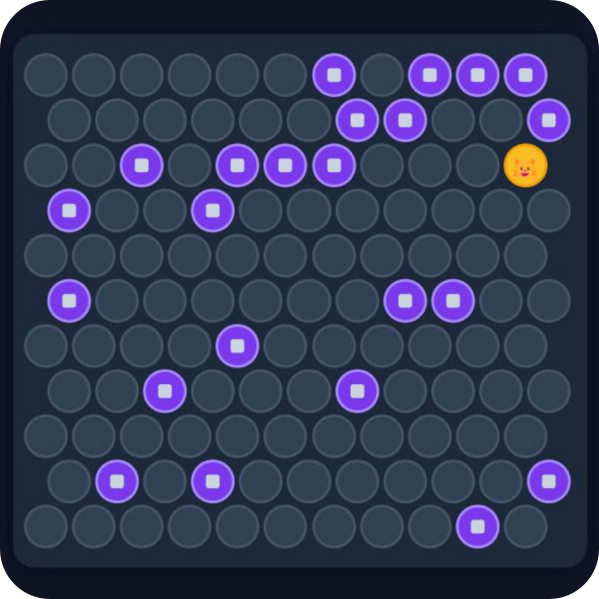}
    &
    \emph{Nearest-cell block}: place a wall on the legal cell closest to the cat.
    This can feel urgent but may leave multiple escape corridors or reachable boundary exits open.
    &
    \textbf{Containment score} based on shortest escape paths, reachable exits, and reachable area.
    The evaluator rewards walls that reshape the global escape structure.
    &
    \textbf{Local blocking $\rightarrow$ global containment.}
    Improvement means placing walls to close bottlenecks, reduce exits, and compress the reachable region.
    \\

    \bottomrule
  \end{tabularx}
  \endgroup
  \caption{Summary of the three game environments. Each row shows the representative interface, the natural greedy baseline, the corresponding global evaluator, and the behavioral shift used to diagnose experience-sensitive learning.}
  \label{tab:environment-unified}
\end{table*}

\section{Related Work}
\paragraph{Game Benchmarks.}
Games have long been used as controlled environments for evaluating language-based decision making.
Early text-game platforms such as TextWorld \citep{cote2018textworld} and Jericho \citep{hausknecht2020jericho} provide standardized environments for training and evaluating agents in text-based interactive worlds.
These environments have enabled a line of work on action generation, exploration, and policy learning in text games, including CALM \citep{yao2020calm}, which trains action language models from human gameplay trajectories.
Recent work has expanded game-based evaluation beyond simple rule-following tasks toward broader interactive, strategic, and multimodal settings.
Several benchmarks evaluate LLM agents in out-of-distribution or competitive games, emphasizing strategic reasoning, long-horizon decision making, and comparisons with human or model opponents \citep{costarelli2024gamebench,guertler2025textarena,tang2026dsgbench}.
Other efforts further broaden the scope to large collections of text-based, visual, and video-game environments, enabling evaluation of interactive gameplay, visual grounding, and multi-turn reasoning across LLMs and VLMs \citep{shi2025korgym,hu2025gamearena,paglieri2025balrog,hu2026lmgamebench,park2026orak}.

\paragraph{Self-Evolving Language Agents.}
Prior work has shown that language models can act as agents in complex sequential environments through reasoning, acting, and search, but these methods mainly improve within-episode deliberation rather than persistent learning from accumulated experience \citep{wei2022cot,yao2023react,yao2023tot,zhou2024lats}.
Another line of work equips agents with memory or reusable experience representations, allowing them to store and reuse reflections, skills, causal abstractions, rules, or procedural knowledge across trials \citep{shinn2023reflexion,wang2023voyager,majumder2023clin}.
Most related to our setting are recent studies of test-time learning and self-evolving agents.
EvoTest formalizes repeated gameplay as J-TTL, where agents are evaluated by improvement across episodes in Jericho games; JitRL and Meta-TTL further study training-free policy adaptation and learnable adaptation policies for agents interacting repeatedly with an environment \citep{he2025evotest,li2026jitrl,lou2026metattl}.
ReasoningBank distills successful and failed experiences into reusable reasoning strategies for test-time self-evolution, while CEL uses repeated gameplay to induce explicit environment rules and a strategic playbook \citep{ouyang2026reasoningbank,wang2025cel}.
These works mainly propose mechanisms for improving agent performance from experience.
%
% In contrast, our contribution is to provide games, human trajectories, and behavioral metrics for identifying when and how repeated experience changes decision-making in interpretable ways.

% \paragraph{Process-Level Evaluation.}
% %
% Many benchmarks evaluate agents using final outcomes such as win rate, task success, or accumulated score.
% %
% Recent work has argued that these metrics can hide important differences in how agents reason, communicate, and act.
% %
% DSGBench includes fine-grained decision dimensions and automated decision tracking for strategic games \citep{tang2026dsgbench}. 
% %
% GameArena analyzes live game sessions to infer fine-grained reasoning capabilities from gameplay traces \citep{hu2025gamearena}. 
% %
% M3-BENCH studies mixed-motive games through process-aware evaluation, including behavioral trajectories, reasoning processes, and communication content \citep{xie2026m3bench}.
% %
% AgentBoard similarly argues for broader evaluation of multi-turn agents beyond simple success rates \citep{chang2024agentboard}. 
% %
% Work on tracing reasoning processes with strategic games also proposes metrics for planning, revision, and resource-aware decision making \citep{yuan2025tracing}.

\paragraph{Process-Level Evaluation and Human Gameplay.}
Many works evaluate agents using final outcomes such as win rate, task success, or accumulated score.
Recent work has argued that such metrics can hide important differences in how agents reason, communicate, and act, and has therefore introduced more fine-grained evaluation protocols based on decision tracking, behavioral trajectories, reasoning traces, planning, revision, and resource-aware decision making \citep{tang2026dsgbench, hu2025gamearena, xie2026m3bench, chang2024agentboard, yuan2025tracing}.
Our work follows this process-level perspective but focuses specifically on whether repeated gameplay produces persistent and interpretable changes.

Human data have also been used in prior game-based NLP work, either as demonstrations for training agents or as reference points for contextualizing model performance \citep{yao2020calm,costarelli2024gamebench,guertler2025textarena,hu2025gamearena}.
However, these works typically treat humans as a source of supervision or as a performance ceiling, rather than as learning subjects whose behavior changes through repeated gameplay.
Our perspective is motivated by classic studies of expertise and cognitive learning, which show that human decision making is shaped by reusable perceptual chunks, templates, heuristics, and abstractions acquired from experience \citep{chase1973perception,gobet1996templates,gobet1998pattern,gershman2015computational,lieder2020resource,gershman2014amortized,dasgupta2018remembrance,dasgupta2020learning}.
We operationalize this view in interactive games: by collecting human trajectories across repeated play, we study not only whether players improve, but also which behavioral patterns change, which misleading heuristics disappear, and which strategies emerge over time.

% \paragraph{Positioning.}
% %
% Our work lies at the intersection of the above areas.
% %
% Compared with broad game benchmarks, we focus on games selected for their potential to reveal experience-driven learning.
% %
% Compared with self-evolving agent methods, we do not propose a new memory or adaptation algorithm; instead, we provide a setting for evaluating whether such mechanisms produce persistent and interpretable strategy changes.
% %
% Compared with process-aware evaluation, we emphasize behavioral evolution across repeated gameplay, not only within-episode reasoning traces.

% Finally, compared with prior human-agent comparisons, we treat human players not merely as a gold standard but as learning subjects whose trajectories provide a reference for experience-sensitive learning.
% %
% Together, our framework turns games from outcome-based tests of agent performance into interactive environments for studying experience-driven behavioral change in both humans and language agents.

\section{Environments and Behavioral Metrics}
\label{sec:environments}
We use games because repeated experience can produce observable changes in action choices.
We select environments with a common diagnostic structure: players may initially favor locally appealing actions, whereas stronger play requires evaluating longer-term strategic consequences.
This allows us to test whether repeated play induces a greedy-to-global shift in decision-making, rather than merely improving final outcomes such as win rate or score.

\subsection{Environment Suite}
\label{subsec:environment-suite}

We study three games: Four in a Row, Othello6, and CircleCat.
They were selected according to four criteria:
\begin{enumerate}[leftmargin=*,noitemsep,topsep=2pt]
    \item success should require reusable strategic principles rather than only local pattern matching;
    \item each game should contain a natural but potentially misleading greedy heuristic;
    \item each game should admit interpretable process-level metrics computable from action traces;
    \item each game should be suitable for both human data collection and agent evaluation under a shared interface.
\end{enumerate}
Under these criteria, the three environments instantiate complementary forms of the same greedy-to-global structure:
Four in a Row tests movement from local line completion toward solver-aligned adversarial search;
Othello6 tests movement from immediate disc maximization toward mobility-aware positional evaluation;
and CircleCat tests movement from proximity-based blocking toward global containment planning.
We summarize these design choices in Table~\ref{tab:environment-unified}.

\paragraph{Four in a Row}
is an adversarial game played on a vertical grid with seven columns.
On each turn, the player chooses a column, and the piece falls to the lowest available cell in that column.
The first player to form four consecutive pieces horizontally, vertically, or diagonally wins; if the board is filled without a winner, the game ends in a draw.

Four in a Row serves as our solver-aligned adversarial search environment.
Its rules are simple enough for repeated human and agent play, but strong play requires more than extending visible lines or blocking immediate threats.
Players must learn to recognize immediate wins, forced threats, multi-step threats, and the strategic value of central columns.
Because legal actions can be evaluated against a strong solver, this environment allows us to measure whether repeated gameplay shifts players from locally salient pattern-based moves toward solver-aligned strategic choices, rather than only improving the final win rate.

\paragraph{Othello6}
is a 6x6 version of Othello.
Players take turns placing discs on the board.
A legal move must bracket at least one line of opponent discs horizontally, vertically, or diagonally, after which the bracketed discs are flipped to the player's color.
If a player has no legal move, the turn is passed.
The game ends when no further moves are available, and the player with more discs wins.

Othello6 serves as our environment for studying greedy-trap avoidance.
It exposes a common form of myopic play: maximizing the number of discs flipped in the current move.
This heuristic is visually salient and often appears reasonable to novice players, but it can reduce future mobility, expose unstable frontier discs, open dangerous positions, or give the opponent access to stronger regions of the board.
The strategic learning target is therefore a shift from immediate flip maximization toward mobility-aware and positionally robust evaluation.
By comparing flip count, mobility, and positional metrics over repeated play, Othello6 allows us to test whether experience changes the heuristic used to evaluate actions.

\begin{table*}[t]
  \centering
  \scriptsize
  \setlength{\tabcolsep}{5pt}
  \renewcommand{\arraystretch}{1.12}
  \begin{tabularx}{\linewidth}{p{0.12\linewidth}p{0.06\linewidth}p{0.18\linewidth}X}
    \toprule
    \textbf{Game} & \textbf{Metric} & \textbf{Strategic Signal} & \textbf{Computation} \\
    \midrule

    Four in a Row
    & \texttt{EAR}
    & Solver alignment.
    & Fraction of player moves that match the solver-best column, measuring overall alignment with globally strong play. \\
    \cmidrule(l){2-4}

    & \texttt{CPQ}
    & Critical-state decision quality.
    & Accuracy on positions where the solver-best action is substantially better than the alternatives, focusing on high-stakes decisions. \\
    \cmidrule(l){2-4}

    & \texttt{OCI}
    & Opening center control.
    & Center preference score over the player's early moves, with more central columns receiving higher scores because they participate in more potential winning lines. \\

    \midrule

    Othello6
    & \texttt{MCI}
    & Mobility control.
    & Difference between the player's future legal-move count and the opponent's future legal-move count after the player's move, measuring whether the player preserves options while restricting the opponent. \\
    \cmidrule(l){2-4}

    & \texttt{CPQ}
    & Search-critical decision quality.
    & Accuracy on search-critical positions where the search-best move is substantially better than the alternatives, capturing decision quality when immediate flip maximization may be misleading. \\

    \midrule

    CircleCat
    & \texttt{EPD}
    & Escape-path lengthening.
    & Increase in the cat's shortest path distance to any boundary cell after the wall is placed, computed from the pre- and post-wall board states. \\
    \cmidrule(l){2-4}

    & \texttt{EPC}
    & Escape-option reduction.
    & Reduction in the number of boundary cells reachable by the cat after the wall is placed, measuring whether the wall removes alternative exits. \\
    \cmidrule(l){2-4}

    & \texttt{CAR}
    & Reachable-region compression.
    & Reduction in the size of the region reachable from the cat's current position after the wall is placed, measuring global containment of the cat's escape area. \\

    \bottomrule
  \end{tabularx}
  \caption{
  Game-specific behavioral metrics used to diagnose how strategic improvement occurs within each environment.
  These metrics complement the cross-game \texttt{GVD} and \texttt{GTA} measures by capturing environment-specific forms of strategic behavior.
  All listed metrics are computed from action traces, and larger values indicate stronger behavior along the corresponding strategic dimension.
  }
  \label{tab:game-specific-metrics}
\end{table*}

\begin{figure*}
    \centering
    \includegraphics[width=\linewidth]{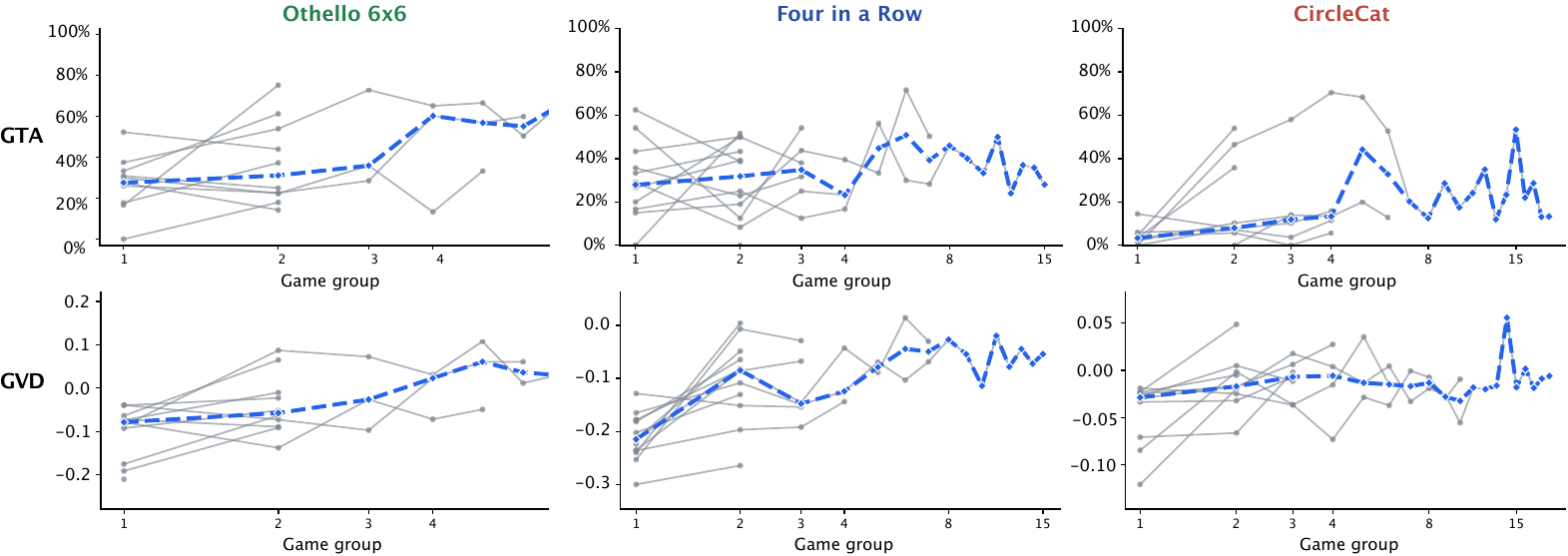}
    \caption{Human trajectories on cross-game greedy-to-global metrics.
Gray lines show individual participants, and the dashed blue line shows the human median across game groups.}
    \label{fig:human_game_greedy}
\end{figure*}

\begin{figure*}[t]
    \centering
    \begin{subfigure}[t]{0.49\textwidth}
        \centering
        \includegraphics[height=0.4\textheight,keepaspectratio]{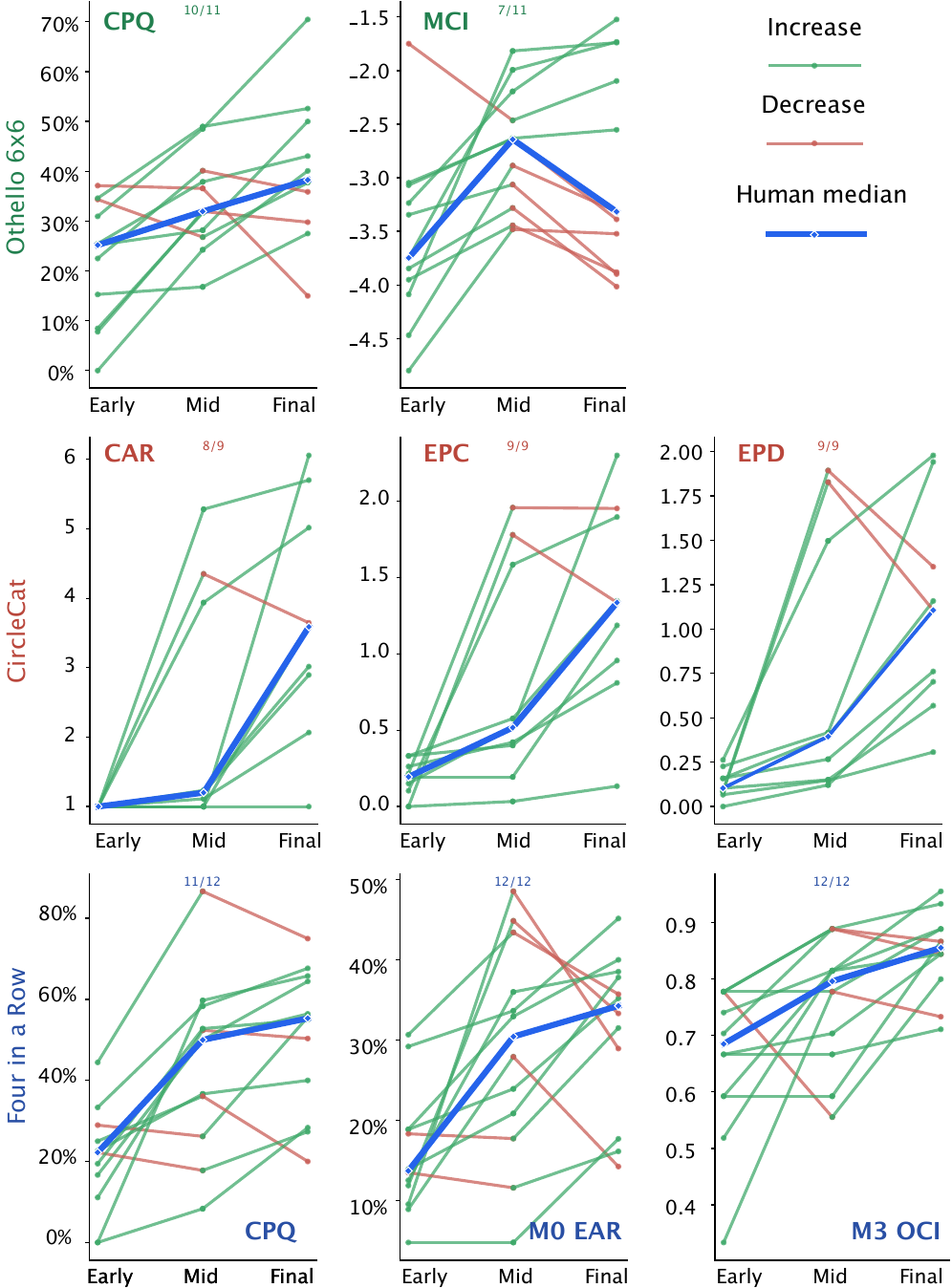}
        \caption{Human trajectories.}
        \label{fig:human_game_sepecific}
    \end{subfigure}
    \hfill
    \begin{subfigure}[t]{0.49\textwidth}
        \centering
        \includegraphics[height=0.4\textheight,keepaspectratio]{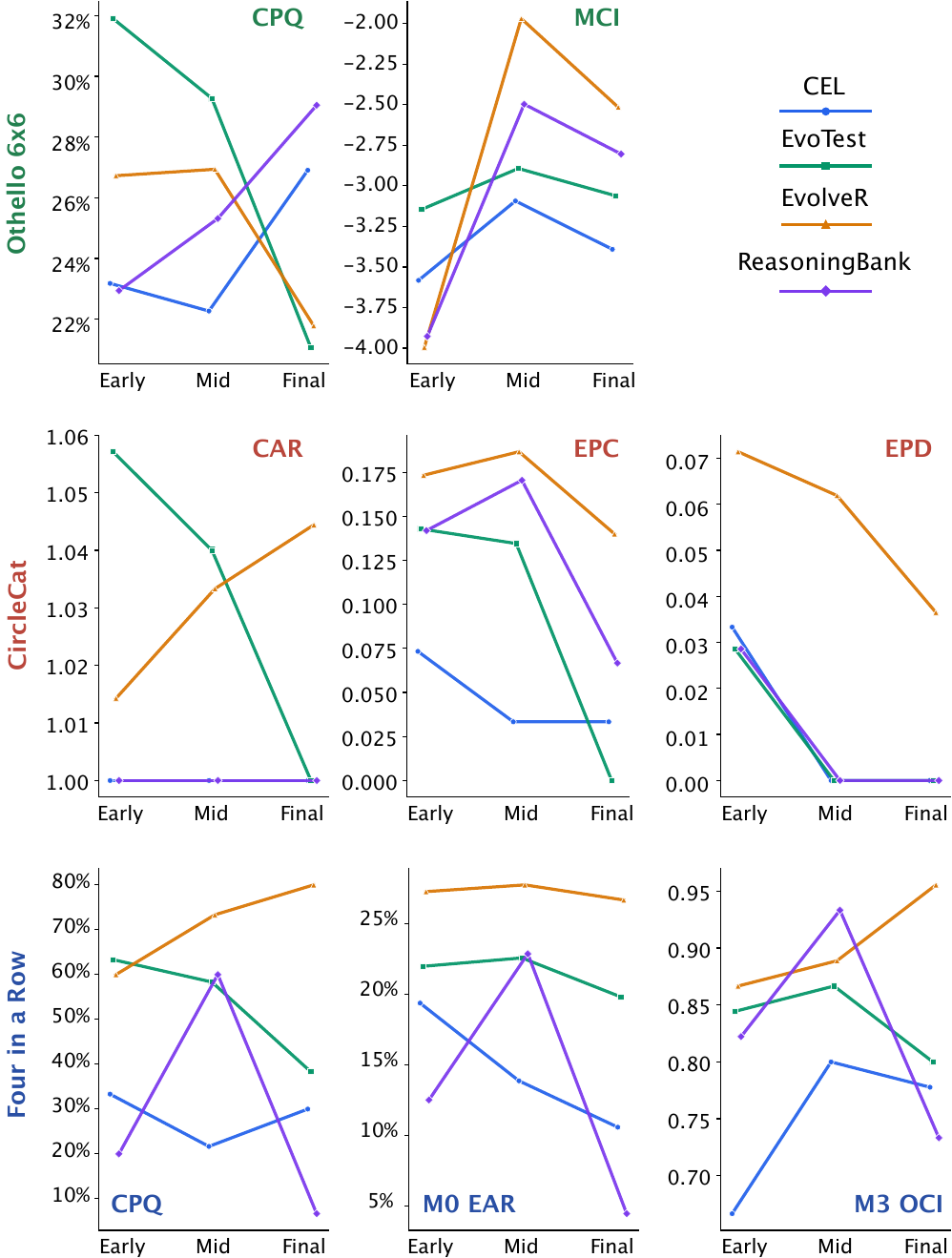}
        \caption{Agent trajectories.}
        \label{fig:agent_game_sepecific}
    \end{subfigure}

    \caption{
    Game-specific behavioral metrics for humans and self-evolving agents.
    The left panel shows human trajectories across Early, Mid, and Final stages, where thin lines denote individual participants and the thick blue line denotes the human median.
    The right panel shows the corresponding trajectories of self-evolving agent methods over the same stages.
    Panel annotations report the fraction of runs or participants whose metric improves.
    }
    \label{fig:game_specific_human_agent}
\end{figure*}

\begin{figure*}
    \centering
    \includegraphics[width=\linewidth]{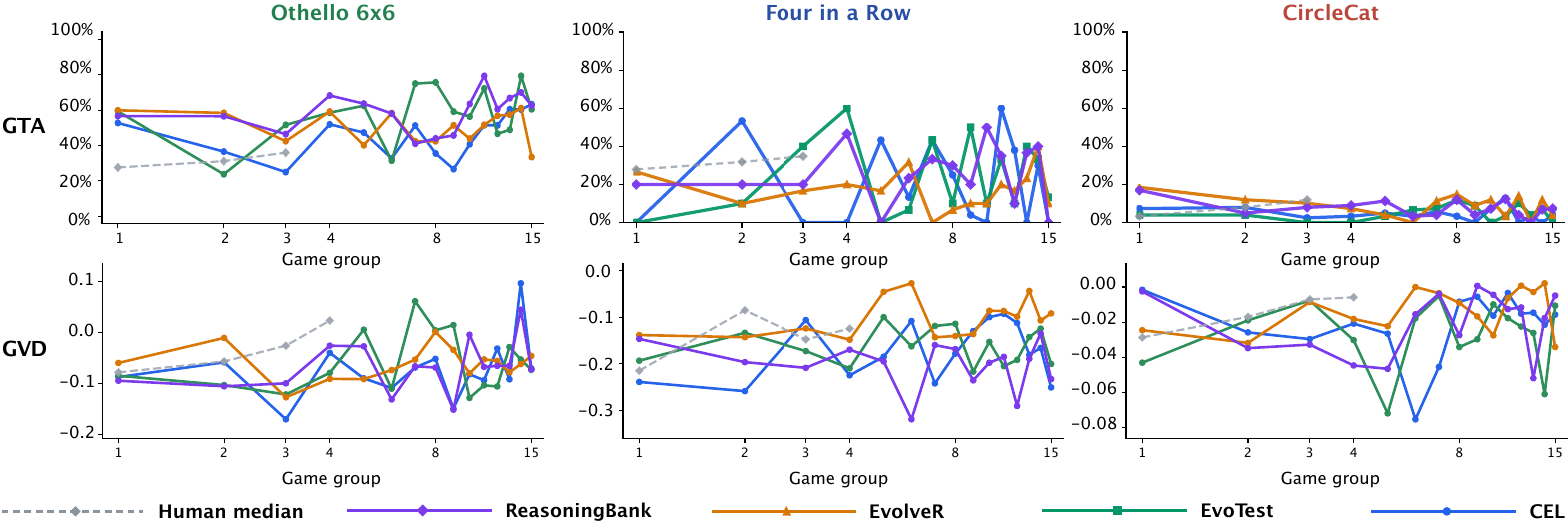}
    \caption{Cross-game greedy-to-global metrics for self-evolving agents. Solid lines show agent trajectories across repeated gameplay, and the dashed gray line shows the human median for reference.}
    \label{fig:agent_game_greedy}
\end{figure*}

\paragraph{CircleCat}
is a spatial containment game on a hex-neighbor board.
On each turn, the player places one wall on an empty cell, after which the cat moves one step.
The player wins if the cat becomes trapped and can no longer reach the boundary, and loses if the cat reaches the boundary.

CircleCat complements the adversarial board games by testing spatial containment rather than opponent modeling.
Its natural novice heuristic is to block cells close to the cat, but effective play requires reasoning about the global escape structure of the board.
A locally urgent wall may leave multiple escape corridors open, whereas a farther wall may close a bottleneck, remove several reachable exits, or shrink the cat's reachable region.

By tracking wall placement relative to the shortest escape paths, reachable boundary cells, and containment regions, CircleCat allows us to test whether repeated play shifts players from local reaction to global containment planning.

\subsection{Metric Design}
\label{subsec:metric-design-principles}
\paragraph{Principles.}
We use process-level metrics to make experience-sensitive behavioral changes observable.
Rather than treating final outcomes such as win rate, score, or game length as the only evidence of improvement, these metrics characterize how players select actions across repeated games.
They are computed from observable actions and can be applied consistently to both human players and language agents.
Together, they connect game-specific behavior to a shared cross-game hypothesis: repeated gameplay should move players away from locally attractive but strategically weak actions and toward globally stronger decisions.

\paragraph{Cross-Game Greedy-to-Global Metrics}
We define two cross-game metrics to quantify whether repeated play moves decisions from local heuristics toward global evaluation.
Greedy Value Difference (\texttt{GVD}) measures the average improvement of a player's actions over a game-specific greedy baseline, while Greedy Trap Avoidance (\texttt{GTA}) focuses on states where that greedy baseline leads to a strategically poor action.

Let $s_t$ denote the game state at decision step $t$, $a_t$ denote the player's action, and $A(s_t)$ denote the legal action set.
For each game, we define a greedy baseline action $g_t \in A(s_t)$ that captures the natural local heuristic in that environment.
We also define a game-specific action evaluator $V(s,a)$, normalized so that larger values indicate actions that are better for the player.
Depending on the environment, $V$ may be derived from a solver, a depth-limited search procedure, or a spatial containment score.

\emph{Greedy Value Difference (\texttt{GVD})}
measures whether the player's action is better than the game-specific greedy baseline.
Because the raw evaluator scale differs across games, we use a softmax-normalized value difference:
\begin{equation}
\begin{aligned}
  \mathrm{GVD}
  &=
  \frac{1}{|\mathcal{T}|}
  \sum_{t \in \mathcal{T}}
  \left[
    P_t(a_t) - P_t(g_t)
  \right], \\
  P_t(a)
  &=
  \frac{\exp(V(s_t,a)/\tau_s)}
  {\sum_{a' \in A(s_t)} \exp(V(s_t,a')/\tau_s)} .
\end{aligned}
\label{eq:gvd}
\end{equation}
Here $V(s_t,a)$ is the game-specific action evaluator, $g_t$ is the locally greedy action, and $\tau_s$ is a temperature parameter.
Higher \texttt{GVD} indicates that the player chooses actions with greater normalized value than the greedy baseline.

\emph{Greedy Trap Avoidance (\texttt{GTA})}
focuses only on states where the greedy baseline is substantially worse than some available alternative.
We define the set of greedy-trap opportunities as
\begin{equation}
  \mathcal{T}_{\mathrm{trap}} =
  \left\{
    t \in \mathcal{T} :
    \max_{a \in A(s_t)} V(s_t, a) - V(s_t, g_t) \geq \tau
  \right\},
  \label{eq:trap-set}
\end{equation}
where $\tau$ is a game-specific threshold for identifying states in which the greedy action is meaningfully suboptimal.
On these states, \texttt{GTA} measures the fraction of times the player avoids the greedy trap by choosing an action that is better than the greedy baseline:
\begin{equation}
  \mathrm{GTA} =
  \frac{1}{|\mathcal{T}_{\mathrm{trap}}|}
  \sum_{t \in \mathcal{T}_{\mathrm{trap}}}
  \mathbf{1}
  \left[
    V(s_t, a_t) - V(s_t, g_t) > 0
  \right].
  \label{eq:gta}
\end{equation}
Thus, \texttt{GVD} captures the magnitude of average improvement over the greedy baseline, whereas \texttt{GTA} captures the frequency with which players avoid clearly harmful greedy choices.

\paragraph{Game-Specific Behavioral Metrics}
\label{subsec:game-specific-metrics}
The cross-game metrics \texttt{GVD} and \texttt{GTA} measure whether players move away from locally greedy actions toward globally stronger decisions.
To further characterize how this shift occurs within each environment, we define a small set of game-specific behavioral metrics targeting the main strategic structure of each game.
These metrics are diagnostic complements rather than separate evaluation goals: they indicate whether improvement is associated with solver alignment, critical-state decision quality, mobility control, or spatial containment.
Table~\ref{tab:game-specific-metrics} summarizes these metrics, their strategic interpretation, how they are computed from action traces, and the preferred direction of change.
Detailed computation procedures for all metrics are provided in the supplementary appendix.

\section{Results}

\subsection{Human Gameplay}
\label{subsec:human-results}

\paragraph{Participants and protocol.}
We collected human gameplay trajectories for the three environments from $32$ participants, including master and doctoral students, resulting in a total of $709$ games.
After filtering incomplete or invalid sessions, the retained data consist of $12$ participants for Four in a Row, $11$ for Othello6, and $9$ for CircleCat.
To reduce the influence of pre-existing game-specific strategies, participants were required to report no recent experience with the corresponding games.
Each participant completed at least $11$ full games.
For analysis, we group consecutive games into five-game bins.
We divide each trajectory into Early, Mid, and Final stages, corresponding to the first three games, the middle five games, and the final three games, respectively; when participants played additional games, later bins are included in the full learning curves.

\paragraph{Cross-game greedy-to-global trends.}
Figure~\ref{fig:human_game_greedy} shows the human trajectories for the cross-game greedy-to-global metrics, \texttt{GTA} and \texttt{GVD}.
Across the three environments, the median human trajectory improves within the first two game groups, suggesting that participants begin to move away from locally greedy heuristics after limited gameplay experience.
The form of this shift differs by environment: in Othello6, improvement corresponds to avoiding flip-greedy traps; in Four in a Row, it corresponds to selecting moves with stronger global or solver value; and in CircleCat, the noisier trajectory still suggests an early movement from proximity-based blocking toward more global containment decisions.
Thus, the cross-game metrics indicate that repeated play changes how participants evaluate actions, rather than only making them more familiar with the interface.

\paragraph{Game-specific behavioral changes.}
Figure~\ref{fig:human_game_sepecific} further examines whether this greedy-to-global shift is accompanied by interpretable changes in game-specific strategic behavior.
The results show broad improvement along the strategic dimensions targeted by each environment.
In Othello6, \texttt{CPQ} improves for $10/11$ participants and \texttt{MCI} improves for $7/11$, indicating gains in search-critical decisions and mobility-aware play.
In CircleCat, containment metrics improve for most or all participants: \texttt{CAR} improves for $8/9$, while \texttt{EPC} and \texttt{EPD} improve for $9/9$, showing that repeated play helps participants reduce reachable space and escape options.
In Four in a Row, \texttt{CPQ} improves for $11/12$ participants, while \texttt{EAR} and \texttt{OCI} improve for $12/12$, suggesting stronger solver alignment and increased early center control.
Together, these results show that human learning is visible not only in cross-game greedy-to-global metrics, but also in the environment-specific mechanisms through which better play emerges.

\paragraph{Takeaway.}
The human data provide a behavioral reference for experience-sensitive game learning.
Human players tend to move from simple local heuristics toward more global decision criteria over repeated play, and this shift is visible in both cross-game and game-specific metrics.
This provides a target for adaptive language agents: an agent should not only improve final performance, but also exhibit durable and interpretable changes in decision-making behavior.

\subsection{Evaluating Self-Evolving Agents}
\label{subsec:self-evolving-agent-results}

We next evaluate whether recent self-evolving language agents exhibit the experience-sensitive behavioral changes observed in human gameplay.
We consider four representative agent designs: \textsc{CEL} \citep{wang2025cel}, \textsc{EvoTest} \citep{he2025evotest}, \textsc{EvolveR} \citep{wu2025evolver}, and \textsc{ReasoningBank} \citep{ouyang2026reasoningbank}.
For each agent, we run repeated gameplay in Four in a Row, Othello6, and CircleCat, collecting $128$ gameplay episodes under the same logging and evaluation protocol.
We compute both the cross-game greedy-to-global metrics, \texttt{GVD} and \texttt{GTA}, and the game-specific behavioral metrics in Table~\ref{tab:game-specific-metrics}.
Figures~\ref{fig:agent_game_greedy} and~\ref{fig:agent_game_sepecific} summarize the resulting trajectories, with the human median included only as a behavioral reference.

Across the three games, current self-evolving agents do not exhibit the stable behavioral improvement observed in human trajectories.
Although individual methods sometimes improve over short stretches, the gains are often noisy, method-specific, and not consistently maintained across later episodes.
This instability appears in both types of metrics: \texttt{GVD} and \texttt{GTA} show that agents only intermittently move away from locally greedy actions, while the game-specific metrics show that improvements in solver alignment, mobility control, critical-state decisions, or spatial containment can regress after temporary gains.
Thus, the issue is not simply that agents achieve lower final performance than humans, but that their behavioral changes are less durable across repeated gameplay.
The evaluated agents sometimes discover better actions, but they do not reliably consolidate these discoveries into stable decision policies.
In contrast to the more consistent greedy-to-global shift observed in human gameplay, current self-evolution mechanisms remain limited in their ability to turn gameplay experience into persistent strategic change.

\subsection{Why Self-Evolving Agents Fail}

The failures of self-evolving agents suggest that their weaknesses are better understood as a hierarchy of bottlenecks rather than a single failure mode.
We therefore perform a closer qualitative analysis of agent failures and organize the recurring patterns into a four-level hierarchy, ranging from low-level state simulation errors to high-level failures of experience integration.

\paragraph{Bottlenecks in self-evolving gameplay.}
Qualitative traces suggest four recurring bottlenecks that prevent agents from turning experience into stable strategic improvement.
\textbf{Level 1: the simulation bottleneck} concerns whether the agent can reliably compute the next state after an action; for example, in Four in a Row, agents may verbalize immediate wins or blocks but still miscompute gravity, landing positions, or line completion.
\textbf{Level 2: the evaluation bottleneck} concerns whether the agent can compare candidate actions by downstream value, as reflected by metrics such as \texttt{CPQ}, \texttt{MCI}, \texttt{EPD}, \texttt{EPC}, and \texttt{CAR}, which test whether agents move beyond locally attractive actions toward solver-aligned, mobility-aware, or containment-aware decisions.
\textbf{Level 3: the proceduralization bottleneck} concerns whether useful strategic language becomes executable decision-time constraints: agents often store principles such as checking threats, preserving mobility, sealing gateways, or cutting shortest paths, but do not consistently translate them into procedures that filter moves, verify threats, or enforce global constraints.
\textbf{Level 4: the experience-integration bottleneck} concerns whether repeated interaction produces stable behavioral change; even after accumulating substantial memory or reflection text, agents may repeat structurally similar errors, such as tactical mis-simulation in Four in a Row, near-boundary emergency blocking in CircleCat, or endgame mobility collapse in Othello6.

This hierarchy explains why self-evolving agents can sound strategically sophisticated while still failing behaviorally.
A failure at Level 1 corrupts the state information used by later reasoning.
A failure at Level 2 prevents the agent from choosing among legal actions by future value.
A failure at Level 3 prevents verbal strategies from becoming operational policies.
Finally, a failure at Level 4 prevents accumulated experience from producing persistent behavioral adaptation across episodes.
Detailed qualitative traces for each level are provided in the supplementary material, where we further decompose these four bottlenecks into finer failure modes and episode-level case studies.

Taken together, these failures suggest that current self-evolving agents learn the vocabulary of strategy more readily than the executable procedures required to apply strategy.
Their central limitation is not the absence of reflection, but the failure to convert reflection into reliable, verifiable, and reusable changes in decision-making behavior.

\section{Conclusion}

We introduced experience-sensitive game learning as a framework for measuring how humans and language agents change their decision-making through repeated gameplay.
Our human trajectories show clear and interpretable behavioral changes: players increasingly avoid greedy traps and improve along strategic dimensions such as solver alignment, mobility control, and spatial containment.
In contrast, recent self-evolving language agents show noisier and less durable gains.
Although self-evolving agents often generate plausible strategic reflections, their main bottleneck remains converting gameplay experience into executable, reusable, and persistent behavioral change across episodes.
Overall, our environments and metrics provide a shared setting for comparing human learning and agent self-evolution through observable decision-making behavior.

% \clearpage

\section*{Limitations}

\paragraph{Limited number of environments.}
A primary limitation of this work is that our detailed analysis is conducted on three game environments. Although this number is modest, it reflects a deliberate choice rather than an attempt to cover games broadly. In the early stage of this project, we experimented with a wider range of games, including 2048, Merge Fall, Four in a Row, Othello, Tic-Tac-Toe, Sudoku, CrossNumber, and Nuts and Bolts. However, we found that only a subset of games satisfied the requirements of the present study: they needed to support repeated play, expose reusable strategic structure, produce interpretable action traces, and admit process-level behavioral metrics beyond final score or win rate.

Some games were too difficult for the type of experience-sensitive analysis targeted in this paper. For example, Go and Chess require substantial domain knowledge and long-term study, making them unsuitable for observing meaningful learning from a small number of repeated games, both for current agents and for many human participants. Other games, such as 2048 and Merge Fall, involve long-horizon action sequences in which a single episode may contain hundreds of operations. Although such games are interesting and we have collected preliminary data for them, their feedback signals are sparse and delayed, making it difficult to attribute behavioral changes to specific strategic shifts at the current stage.

Therefore, the limited number of environments remains an important limitation. Our current benchmark prioritizes validated environments with clear behavioral metrics over broad game coverage. We are actively extending the evaluation protocol and benchmark to include more verified environments, with the goal of supporting a broader study of self-evolving agents. In this paper, however, we intentionally favor quality over quantity, because the validity and interpretability of each environment are crucial for analyzing experience-sensitive behavioral change.

\paragraph{Variability and cost of human data.}
A second limitation is the difficulty of collecting and interpreting human gameplay data. Human trajectories are valuable because they provide a behavioral reference for experience-sensitive learning, but they are also inherently noisy. Participants differ substantially in attention, prior knowledge, motivation, and learning style. Some participants may become distracted during gameplay, producing trajectories that appear random or inconsistent. Others may learn, but in a direction that does not align with the intended strategic principle of the environment. As a result, human learning curves can contain substantial variability, even when the task interface and instructions are fixed.

We observed such variability in the collected trajectories. Understanding it more deeply will require a larger and more diverse human dataset, as well as more careful modeling of individual differences. However, collecting human gameplay trajectories is costly, especially when the goal is not merely to obtain final scores but to observe repeated-game behavioral change over time. We therefore view the current human study as an initial reference rather than a definitive characterization of human learning. Expanding the scale, diversity, and reliability of human data is an important direction for future work.

\paragraph{Comparison with trainable game agents.}
Another limitation is that we do not compare against reinforcement-learning agents or language models trained directly in the target game environments.
This is a deliberate scope choice.
Reinforcement learning follows a different adaptation paradigm: an agent is optimized for a specific environment through repeated training, often achieving performance far beyond ordinary human players and, in some games, even expert-level play.
However, such environment-specific optimization does not directly reflect the type of experience-sensitive behavioral change studied in this work.
Our goal is not to test whether a trainable policy can eventually master a game, but to examine whether language agents can use limited repeated interaction to revise their decision-making in a way that is behaviorally comparable to human learning.

This distinction is important because large language model agents occupy a different setting from conventional trainable game agents.
They enter the environment with broad prior knowledge, natural-language reasoning, memory, and reflection mechanisms, but without gradient-based training on the specific game.
We therefore focus on whether these agents can convert gameplay experience into reusable strategic behavior through interaction alone.
Comparing this form of interaction-based adaptation with fully trainable reinforcement-learning systems is an important direction for future work, but it requires a different experimental protocol and should be interpreted separately from the experience-sensitive learning setting studied here.

\section*{Ethical Considerations}

This work involves the collection and analysis of human gameplay trajectories.
Participants were informed about the purpose of the study, the nature of the gameplay tasks, and the types of data collected before participation.
Participation was voluntary, and participants could stop playing at any time.
The collected data consist of gameplay interaction traces, such as game states, actions, and episode-level outcomes.
We did not collect sensitive personal attributes, and our analysis focuses on aggregate behavioral patterns rather than individual performance evaluation.
The study procedure was reviewed and approved by the project supervisor through the available internal project oversight process before data collection.

To protect participant privacy, human gameplay records were anonymized before analysis.
Any identifiers used during data collection were removed or replaced with anonymous participant IDs.
The reported results are aggregated across participants, and no individual participant is identifiable from the figures, tables, or qualitative analysis.
The tasks used in this study are simple puzzle or board-game environments and involve minimal risk beyond ordinary interaction with computer games.

Because human gameplay can vary substantially across individuals, we use the data as a behavioral reference for studying experience-sensitive learning rather than as a normative judgment of participant ability.
We do not use the data to make claims about participants' cognitive capacity, intelligence, or personal traits.
If required by the venue or institution, the collection protocol should be reviewed under the applicable human-subjects research guidelines.

% \section*{Acknowledgments}

\bibliography{custom}

@InProceedings{cote2018textworld,
author="C{\^o}t{\'e}, Marc-Alexandre
and K{\'a}d{\'a}r, {\'A}kos
and Yuan, Xingdi
and Kybartas, Ben
and Barnes, Tavian
and Fine, Emery
and Moore, James
and Hausknecht, Matthew
and El Asri, Layla
and Adada, Mahmoud
and Tay, Wendy
and Trischler, Adam",
editor="Cazenave, Tristan
and Saffidine, Abdallah
and Sturtevant, Nathan",
title="TextWorld: A Learning Environment for Text-Based Games",
booktitle="Computer Games",
year="2019",
publisher="Springer International Publishing",
address="Cham",
pages="41--75",
url="https://doi.org/10.1007/978-3-030-24337-1_3",
doi="10.1007/978-3-030-24337-1_3",
isbn="978-3-030-24337-1"
}

@article{hausknecht2020jericho,
title={Interactive Fiction Games: A Colossal Adventure}, 
volume={34}, 
url={https://ojs.aaai.org/index.php/AAAI/article/view/6297}, 
doi={10.1609/aaai.v34i05.6297}, 
number={05}, 
journal={Proceedings of the AAAI Conference on Artificial Intelligence}, 
author={Hausknecht, Matthew and Ammanabrolu, Prithviraj and Côté, Marc-Alexandre and Yuan, Xingdi}, 
year={2020}, 
month={Apr.},
pages={7903--7910}  }

@inproceedings{yao2020calm,
    title = "Keep {CALM} and Explore: Language Models for Action Generation in Text-based Games",
    author = "Yao, Shunyu  and
      Rao, Rohan  and
      Hausknecht, Matthew  and
      Narasimhan, Karthik",
    editor = "Webber, Bonnie  and
      Cohn, Trevor  and
      He, Yulan  and
      Liu, Yang",
    booktitle = "Proceedings of the 2020 Conference on Empirical Methods in Natural Language Processing (EMNLP)",
    month = nov,
    year = "2020",
    address = "Online",
    publisher = "Association for Computational Linguistics",
    url = "https://aclanthology.org/2020.emnlp-main.704/",
    doi = "10.18653/v1/2020.emnlp-main.704",
    pages = "8736--8754"
}

@article{costarelli2024gamebench,
  title={{Gamebench}: Evaluating strategic reasoning abilities of llm agents},
  author={Costarelli, Anthony and Allen, Mat and Hauksson, Roman and Sodunke, Grace and Hariharan, Suhas and Cheng, Carlson and Li, Wenjie and Clymer, Joshua and Yadav, Arjun},
  journal={arXiv preprint arXiv:2406.06613},
  url={https://arxiv.org/abs/2406.06613},
  year={2024}
}

@article{guertler2025textarena,
  title={Textarena},
  author={Guertler, Leon and Cheng, Bobby and Yu, Simon and Liu, Bo and Choshen, Leshem and Tan, Cheston},
  journal={arXiv preprint arXiv:2504.11442},
  year={2025},
  url={https://arxiv.org/abs/2504.11442}
}

@inproceedings{tang2026dsgbench,
  title={Dsgbench: A diverse strategic game benchmark for evaluating llm-based agents in complex decision-making environments},
  author={Tang, Wenjie and Zhou, Yuan and Cheng, Keyan and Xu, Erqiang and Xiao, Liquan and Li, Minne},
  booktitle={ICASSP 2026 - 2026 IEEE International Conference on Acoustics, Speech and Signal Processing (ICASSP)},
  pages={16987--16991},
  year={2026},
  organization={IEEE},
  doi={10.1109/ICASSP55912.2026.11460704},
  url={https://ieeexplore.ieee.org/document/11460704}
}

@inproceedings{shi2025korgym,
 author = {Shi, Jiajun and Yang, Jian and Liu, Jiaheng and Bu, Xingyuan and Chen, Jiangjie and Zhou, Junting and Ma, Kaijing and Wen, Zhoufutu and Wang, Bingli and He, Yancheng and Song, Liang and Zhu, Hualei and Li, Shilong and Wang, Xingjian and Zhang, Wei and Yuan, Ruibin and Yao, Yifan and Yang, Wenjun and Wang, Yunli and Fang, Siyuan and Yuan, Siyu and He, Qianyu and Tang, Robert and Tan, Yingshui and Zhou, Wangchunshu and Zhang, Zhao-Xiang and Li, Zhoujun and Huang, Wenhao and Zhang, Ge},
 booktitle = {Advances in Neural Information Processing Systems},
 editor = {D. Belgrave and C. Zhang and H. Lin and R. Pascanu and P. Koniusz and M. Ghassemi and N. Chen},
 pages = {161286--161314},
 publisher = {Curran Associates, Inc.},
 title = {{KORGym}: A Dynamic Game Platform for LLM Reasoning Evaluation},
 url = {https://proceedings.neurips.cc/paper_files/paper/2025/file/ebfa4297cd6419f64efe86f657ba49d0-Paper-Conference.pdf},
 volume = {38},
 year = {2025}
}

@inproceedings{hu2025gamearena,
 author = {Hu, Lanxiang and Li, Qiyu and Xie, Anze and Jiang, Nan and Stoica, Ion and Jin, Haojian and Zhang, Hao},
 booktitle = {International Conference on Learning Representations},
 editor = {Y. Yue and A. Garg and N. Peng and F. Sha and R. Yu},
 pages = {33278--33309},
 title = {GameArena: Evaluating {LLM} Reasoning through Live Computer Games},
 url = {https://proceedings.iclr.cc/paper_files/paper/2025/file/520416e27d3b0cef3cd70a083e2991c7-Paper-Conference.pdf},
 volume = {2025},
 year = {2025}
}

@inproceedings{paglieri2025balrog,
 author = {Paglieri, Davide and Cupia{\l}, Bart{\l} omiej and Coward, Samuel and Piterbarg, Ulyana and Wo{\l} czyk, Maciej and Khan, Akbir and Pignatelli, Eduardo and Kuci{\'n}ski, {\L} ukasz and Pinto, Lerrel and Fergus, Rob and Foerster, Jakob and Parker-Holder, Jack and Rocktaeschel, Tim},
 booktitle = {International Conference on Learning Representations},
 editor = {Y. Yue and A. Garg and N. Peng and F. Sha and R. Yu},
 pages = {96666--96702},
 title = {{BALROG}: Benchmarking Agentic {LLM} and {VLM} Reasoning On Games},
 url = {https://proceedings.iclr.cc/paper_files/paper/2025/file/f0b1515be276f6ba82b4f2b25e50bef0-Paper-Conference.pdf},
 volume = {2025},
 year = {2025}
}

@inproceedings{hu2026lmgamebench,
title={lmgame-Bench: How Good are {LLM}s at Playing Games?},
author={Lanxiang Hu and Mingjia Huo and Yuxuan Zhang and Haoyang Yu and Eric P. Xing and Ion Stoica and Tajana Rosing and Haojian Jin and Hao Zhang},
booktitle={The Fourteenth International Conference on Learning Representations},
year={2026},
url={https://openreview.net/forum?id=qeziG97WUZ}
}

@inproceedings{park2026orak,
title={Orak: A Foundational Benchmark for Training and Evaluating {LLM} Agents on Diverse Video Games},
author={Dongmin Park and Minkyu Kim and Beongjun Choi and Junhyuck Kim and Keon Lee and Jonghyun Lee and Inkyu Park and Byeong-Uk Lee and Jaeyoung Hwang and Jaewoo Ahn and Ameya Sunil Mahabaleshwarkar and Bilal Kartal and Pritam Biswas and Yoshi Suhara and Kangwook Lee and Jaewoong Cho},
booktitle={The Fourteenth International Conference on Learning Representations},
year={2026},
url={https://openreview.net/forum?id=H1ncX6O6Yh}
}

@inproceedings{wei2022cot,
 author = {Wei, Jason and Wang, Xuezhi and Schuurmans, Dale and Bosma, Maarten and ichter, brian and Xia, Fei and Chi, Ed and Le, Quoc V and Zhou, Denny},
 booktitle = {Advances in Neural Information Processing Systems},
 editor = {S. Koyejo and S. Mohamed and A. Agarwal and D. Belgrave and K. Cho and A. Oh},
 pages = {24824--24837},
 publisher = {Curran Associates, Inc.},
 title = {Chain-of-Thought Prompting Elicits Reasoning in Large Language Models},
 url = {https://proceedings.neurips.cc/paper_files/paper/2022/file/9d5609613524ecf4f15af0f7b31abca4-Paper-Conference.pdf},
 volume = {35},
 year = {2022}
}

@inproceedings{yao2023react,
title= {{ReAct}: Synergizing Reasoning and Acting in Language Models},
author={Shunyu Yao and Jeffrey Zhao and Dian Yu and Nan Du and Izhak Shafran and Karthik R Narasimhan and Yuan Cao},
booktitle={The Eleventh International Conference on Learning Representations },
year={2023},
url={https://openreview.net/forum?id=WE_vluYUL-X}
}

@inproceedings{yao2023tot,
 author = {Yao, Shunyu and Yu, Dian and Zhao, Jeffrey and Shafran, Izhak and Griffiths, Tom and Cao, Yuan and Narasimhan, Karthik},
 booktitle = {Advances in Neural Information Processing Systems},
 editor = {A. Oh and T. Naumann and A. Globerson and K. Saenko and M. Hardt and S. Levine},
 pages = {11809--11822},
 publisher = {Curran Associates, Inc.},
 title = {Tree of Thoughts: Deliberate Problem Solving with Large Language Models},
 url = {https://proceedings.neurips.cc/paper_files/paper/2023/file/271db9922b8d1f4dd7aaef84ed5ac703-Paper-Conference.pdf},
 volume = {36},
 year = {2023}
}

@inproceedings{zhou2024lats,
author = {Zhou, Andy and Yan, Kai and Shlapentokh-Rothman, Michal and Wang, Haohan and Wang, Yu-Xiong},
title = {Language agent tree search unifies reasoning, acting, and planning in language models},
year = {2024},
publisher = {JMLR.org},
booktitle = {Proceedings of the 41st International Conference on Machine Learning},
articleno = {2572},
numpages = {23},
location = {Vienna, Austria},
url={https://openreview.net/forum?id=njwv9BsGHF}
}

@inproceedings{shinn2023reflexion,
 author = {Shinn, Noah and Cassano, Federico and Gopinath, Ashwin and Narasimhan, Karthik and Yao, Shunyu},
 booktitle = {Advances in Neural Information Processing Systems},
 editor = {A. Oh and T. Naumann and A. Globerson and K. Saenko and M. Hardt and S. Levine},
 pages = {8634--8652},
 publisher = {Curran Associates, Inc.},
 title = {Reflexion: language agents with verbal reinforcement learning},
 url = {https://proceedings.neurips.cc/paper_files/paper/2023/file/1b44b878bb782e6954cd888628510e90-Paper-Conference.pdf},
 volume = {36},
 year = {2023}
}

@article{wang2023voyager,
title={Voyager: An Open-Ended Embodied Agent with Large Language Models},
author={Guanzhi Wang and Yuqi Xie and Yunfan Jiang and Ajay Mandlekar and Chaowei Xiao and Yuke Zhu and Linxi Fan and Anima Anandkumar},
journal={Transactions on Machine Learning Research},
issn={2835-8856},
year={2024},
url={https://openreview.net/forum?id=ehfRiF0R3a}
}

@inproceedings{majumder2023clin,
title={{CLIN}: A Continually Learning Language Agent for Rapid Task Adaptation and Generalization},
author={Bodhisattwa Prasad Majumder and Bhavana Dalvi Mishra and Peter Jansen and Oyvind Tafjord and Niket Tandon and Li Zhang and Chris Callison-Burch and Peter Clark},
booktitle={First Conference on Language Modeling},
year={2024},
url={https://openreview.net/forum?id=xS6zx1aBI9}
}

@inproceedings{he2025evotest,
title={EvoTest: Evolutionary Test-Time Learning for Self-Improving Agentic Systems},
author={Yufei He and Juncheng Liu and Yue Liu and Yibo Li and Tri Cao and Zhiyuan Hu and Xinxing Xu and Bryan Hooi},
booktitle={The Fourteenth International Conference on Learning Representations},
year={2026},
url={https://openreview.net/forum?id=JFnnajbkvP}
}

@article{li2026jitrl,
  title={Just-In-Time Reinforcement Learning: Continual Learning in LLM Agents Without Gradient Updates},
  author={Li, Yibo and Lin, Zijie and Deng, Ailin and Zhang, Xuan and He, Yufei and Ji, Shuo and Cao, Tri and Hooi, Bryan},
  journal={arXiv preprint arXiv:2601.18510},
  year={2026},
  url={https://arxiv.org/abs/2601.18510}
}

@article{lou2026metattl,
  title={Learning to Learn-at-Test-Time: Language Agents with Learnable Adaptation Policies},
  author={Lou, Zhanzhi and Chen, Hui and Li, Yibo and Wang, Qian and Hooi, Bryan},
  journal={arXiv preprint arXiv:2604.00830},
  year={2026},
  url={https://arxiv.org/abs/2604.00830}, 
}

@inproceedings{ouyang2026reasoningbank,
title={ReasoningBank: Scaling Agent Self-Evolving with Reasoning Memory},
author={Siru Ouyang and Jun Yan and I-Hung Hsu and Yanfei Chen and Ke Jiang and Zifeng Wang and Rujun Han and Long Le and Samira Daruki and Xiangru Tang and Vishy Tirumalashetty and George Lee and Mahsan Rofouei and Hangfei Lin and Jiawei Han and Chen-Yu Lee and Tomas Pfister},
booktitle={The Fourteenth International Conference on Learning Representations},
year={2026},
url={https://openreview.net/forum?id=jL7fwchScm}
}

@article{wang2025cel,
  title={Cogito, Ergo Ludo: An Agent that Learns to Play by Reasoning and Planning},
  author={Wang, Sai and Wu, Yu and Xu, Zhongwen},
  journal={arXiv preprint arXiv:2509.25052},
  year={2025},
  url={https://arxiv.org/abs/2509.25052},
}

@inproceedings{chang2024agentboard,
 author = {Ma, Chang and Zhang, Junlei and Zhu, Zhihao and Yang, Cheng and Yang, Yujiu and Jin, Yaohui and Lan, Zhenzhong and Kong, Lingpeng and He, Junxian},
 booktitle = {Advances in Neural Information Processing Systems},
 doi = {10.52202/079017-2365},
 editor = {A. Globerson and L. Mackey and D. Belgrave and A. Fan and U. Paquet and J. Tomczak and C. Zhang},
 pages = {74325--74362},
 publisher = {Curran Associates, Inc.},
 title = {AgentBoard: An Analytical Evaluation Board of Multi-turn LLM Agents},
 url = {https://proceedings.neurips.cc/paper_files/paper/2024/file/877b40688e330a0e2a3fc24084208dfa-Paper-Datasets_and_Benchmarks_Track.pdf},
 volume = {37},
 year = {2024}
}

@article{xie2026m3bench,
  title={M3-BENCH: Process-Aware Evaluation of LLM Agents Social Behaviors in Mixed-Motive Games},
  author={Xie, Sixiong and Shi, Zhuofan and Shen, Haiyang and Huang, Gang and Ma, Yun and Jing, Xiang},
  journal={arXiv preprint arXiv:2601.08462},
  year={2026},
  url={https://arxiv.org/abs/2601.08462}
}

@article{yuan2025tracing,
  title={Tracing llm reasoning processes with strategic games: A framework for planning, revision, and resource-constrained decision making},
  author={Yuan, Xiaopeng and Zhang, Xingjian and Xu, Ke and Xu, Yifan and Yu, Lijun and Wang, Jindong and Dong, Yushun and Wang, Haohan},
  journal={arXiv preprint arXiv:2506.12012},
  year={2025},
  url={https://arxiv.org/abs/2506.12012}
  }

@article{chase1973perception,
  title={Perception in chess},
  author={Chase, William G and Simon, Herbert A},
  journal={Cognitive psychology},
  volume={4},
  number={1},
  pages={55--81},
  year={1973},
  publisher={Elsevier},
  issn = {0010-0285},
  doi = {10.1016/0010-0285(73)90004-2},
  url = {https://doi.org/10.1016/0010-0285(73)90004-2}
}

@article{gobet1996templates,
  title={Templates in Chess Memory: A Mechanism for Recalling Several Boards},
  author={Gobet, Fernand and Simon, Herbert A},
  journal={Cognitive Psychology},
  volume={31},
  number={1},
  pages={1--40},
  year={1996},
  publisher={Elsevier},
  issn = {0010-0285},
  doi = {https://doi.org/10.1006/cogp.1996.0011},
  url = {https://www.sciencedirect.com/science/article/pii/S0010028596900110}
}

@article{gobet1998pattern,
  title={Pattern recognition makes search possible: Comments on Holding (1992)},
  author={Fernand R. Gobet and Herbert A. Simon},
  journal={Psychological Research},
  year={1998},
  volume={61},
  pages={204-208},
  doi = {10.1007/s004260050025},
  url={https://doi.org/10.1007/s004260050025}
}

@article{gershman2015computational,
author = {Samuel J. Gershman  and Eric J. Horvitz  and Joshua B. Tenenbaum },
title = {Computational rationality: A converging paradigm for intelligence in brains, minds, and machines},
journal = {Science},
volume = {349},
number = {6245},
pages = {273-278},
year = {2015},
doi = {10.1126/science.aac6076},
URL = {https://www.science.org/doi/abs/10.1126/science.aac6076}
}

@article{lieder2020resource, 
  title={Resource-rational analysis: Understanding human cognition as the optimal use of limited computational resources}, 
  volume={43}, 
  doi={10.1017/S0140525X1900061X}, 
  journal={Behavioral and Brain Sciences}, 
  author={Lieder, Falk and Griffiths, Thomas L.},
  year={2020},
  url={https://doi.org/10.1017/S0140525X1900061X},
  pages={e1}}

@inproceedings{gershman2014amortized,
  author       = {Samuel Gershman and
                  Noah D. Goodman},
  editor       = {Paul Bello and
                  Marcello Guarini and
                  Marjorie McShane and
                  Brian Scassellati},
  title        = {Amortized Inference in Probabilistic Reasoning},
  booktitle    = {Proceedings of the 36th Annual Meeting of the Cognitive Science Society, CogSci 2014, Quebec City, Canada, July 23-26, 2014},
  publisher    = {cognitivesciencesociety.org},
  year         = {2014},
  url          = {https://escholarship.org/uc/item/34j1h7k5},
  bibsource    = {dblp computer science bibliography, https://dblp.org}
}

@article{dasgupta2018remembrance,
title = {Remembrance of inferences past: Amortization in human hypothesis generation},
journal = {Cognition},
volume = {178},
pages = {67-81},
year = {2018},
issn = {0010-0277},
doi = {https://doi.org/10.1016/j.cognition.2018.04.017},
url = {https://www.sciencedirect.com/science/article/pii/S0010027718301094},
author = {Ishita Dasgupta and Eric Schulz and Noah D. Goodman and Samuel J. Gershman}
}

@article{dasgupta2020learning,
  title={A theory of learning to infer.},
  author={Dasgupta, Ishita and Schulz, Eric and Tenenbaum, Joshua B and Gershman, Samuel J},
  journal={Psychological review},
  volume={127},
  number={3},
  pages={412},
  year={2020},
  publisher={American Psychological Association},
  doi = {10.1037/rev0000178},
  url={https://doi.org/10.1037/rev0000178}
}

@article{wu2025evolver,
  title={EvolveR: Self-Evolving LLM Agents through an Experience-Driven Lifecycle},
  author={Rong Wu and Xiaoman Wang and Jianbiao Mei and Pinlong Cai and Daocheng Fu and Cheng Yang and Licheng Wen and Xuemeng Yang and Yufan Shen and Yuxin Wang and Botian Shi},
  journal={arXiv preprint arXiv:2510.16079},
  year={2025},
  url={https://arxiv.org/abs/2510.16079}
}

\clearpage

\appendix

\section*{Appendix}

\section{Detailed Failure Analysis of Self-Evolving Agents}
\label{app:failure_analysis}

This appendix provides a qualitative failure analysis of self-evolving agents around later episodes.
We organize the analysis using the four-level bottleneck hierarchy introduced in the main paper.
The four levels are nested.
Level 1 concerns whether the agent can simulate the next state.
Level 2 concerns whether it can evaluate candidate actions by future value.
Level 3 concerns whether verbal strategies become executable action constraints.
Level 4 concerns whether accumulated experience produces persistent behavioral change across episodes.

\begin{table*}[t]
\centering
\small
\setlength{\tabcolsep}{5pt}
\renewcommand{\arraystretch}{1.15}
\begin{tabularx}{\textwidth}{
    p{0.10\textwidth}
    p{0.22\textwidth}
    X
}
\toprule
\textbf{Level} & \textbf{Bottleneck} & \textbf{Detailed failure modes} \\
\midrule
Level 1 &
Simulation bottleneck &
State-transition simulation failure: the agent verbally checks legal moves, immediate wins, blocks, or line completion, but fails to reliably simulate the next state. \\

Level 2 &
Evaluation bottleneck &
Counterfactual search failure; reactive local-salience failure. The agent recognizes that future consequences matter, but still fails to compare candidate actions by their downstream value. \\

Level 3 &
Proceduralization bottleneck &
Verbal strategy without proceduralization; abstraction grounding failure. The agent stores useful strategic principles, but does not reliably convert them into action-time constraints. \\

Level 4 &
Experience-integration bottleneck &
Long-horizon credit assignment failure; memory retrieval and compression failure. The agent accumulates memory and reflection, but does not turn them into persistent experience-sensitive behavioral change. \\
\bottomrule
\end{tabularx}
\caption{
A four-level hierarchy of failure modes in self-evolving agents.
The levels are nested: reliable state simulation supports future-oriented evaluation;
evaluation supports proceduralized strategy execution; and proceduralized execution is required for experience-sensitive behavioral change across episodes.
}
\label{tab:failure_hierarchy}
\end{table*}

\subsection{Level 1: Simulation Bottleneck}

\paragraph{Failure Analysis 1: State-Transition Simulation Failure.}

\textbf{Explanation.}
This failure occurs when an agent knows the game rules in language but cannot reliably simulate the next state after an action.
In board games, this means that the agent may correctly state that it should check legal moves, immediate wins, blocks, gravity, or line completion, yet still miscompute the actual board transition.
This is the lowest-level bottleneck in the hierarchy.
If the agent cannot accurately compute the state after a candidate move, then its later strategic evaluation is built on an unreliable representation of the game.

\textbf{Case: Four in a Row, CEL, Episode 50.}
A clear example appears in Four in a Row, Episode 50, with the CEL agent.
By this point, the agent had already accumulated a very long strategy text, with \texttt{strategy\_before\_len = 29012}.
Nevertheless, the episode ended in \texttt{LOSE} with \texttt{score = bot\_win}.
The failure is therefore not an early-stage failure caused by lack of exposure.
It occurs after substantial accumulated strategy text.

In step 13, the agent selected action \texttt{4} and described it as an immediate winning move.
Its reasoning claimed that the piece would land in column 4, row 2, completing a vertical four across rows 2, 3, 4, and 5.
However, the game did not end with an agent win.
The final result was still \texttt{bot\_win}.
This indicates that the agent's action choice was based on a faulty simulation of the board transition, not merely on a weak high-level strategy.

The same episode also shows instability in tactical scanning.
In step 17, the agent cycled through several possible immediate wins and bot threats.
It first considered column 1 as a possible immediate winning move, then rejected it.
It then checked whether the bot had an immediate win in column 5, rejected that as well, and eventually identified column 3 as a possible bot winning threat.
This pattern shows that the agent knows the correct verbal routine, namely to scan immediate wins and blocks, but the scan itself is unstable.

\textbf{Interpretation.}
This case shows that late-stage self-evolving agents may possess the language of tactical reasoning without a reliable internal state simulator.
The agent can say that it is checking immediate threats, but it does not consistently compute piece placement, gravity, or line completion correctly.
In the four-level hierarchy, this is a Level 1 failure because it precedes strategic evaluation.
If the agent cannot trust its own next-state computation, then solver alignment and critical-state decision quality cannot be stable.

\subsection{Level 2: Evaluation Bottleneck}

\paragraph{Failure Analysis 2: Counterfactual Search Failure.}

\textbf{Explanation.}
This failure occurs when the agent recognizes that multiple candidate actions must be compared, but does not reliably evaluate their downstream consequences.
The agent may say that a state requires concrete sequence analysis, minimax-like reasoning, or future mobility evaluation.
However, its chosen action does not reflect a stable comparison among possible future trajectories.
This is a Level 2 failure because the agent's problem is not only computing the next state, but assigning value to different possible futures.

This bottleneck is directly connected to our process-level metrics.
Four in a Row uses solver alignment and critical-state decision quality to test whether the agent chooses globally strong columns.
Othello6 uses mobility control and search-critical decision quality to test whether the agent avoids moves that look good immediately but damage future options.
CircleCat uses escape-path lengthening, escape-option reduction, and reachable-region compression to test whether the agent evaluates walls by their effect on the global escape structure.

\textbf{Case: Othello6, ReasoningBank, Episode 50.}
In Othello6, Episode 50, the ReasoningBank agent entered the episode with substantial memory and strategy text.
The metadata reports \texttt{strategy\_before\_len = 15199}, yet the final score was \texttt{12:24}, a large loss.
The agent's memory already included a corner-adjacency audit, instructing it to test each open corner for the opponent's next move using the exact bracket rule.

The most revealing moment occurs in step 27.
The agent explicitly recognized that it was in a high-importance endgame state and that only two legal moves remained.
It stated that the best action should be chosen through concrete sequence analysis rather than broad heuristics such as central control or edge shape.
This is the right verbal diagnosis of the situation.
However, the later trajectory still produced forced passes in step 31 and step 33, where Black had no legal move and had to pass.

\textbf{Interpretation.}
The agent can name the required evaluation procedure, but it does not execute that procedure reliably.
If it had successfully compared the legal moves by their future mobility consequences, it should have been able to anticipate which move would preserve legal options and which move would collapse into forced passes.
The failure therefore lies in counterfactual evaluation: the agent describes future-sensitive reasoning, but its action does not reflect a correct comparison of future states.

\paragraph{Failure Analysis 3: Reactive Local-Salience Failure.}

\textbf{Explanation.}
This failure occurs when the agent remains driven by the most locally urgent or visually salient threat, rather than evaluating the global structure of the game.
It is especially clear in CircleCat.
The natural novice heuristic is to place a wall near the cat, but effective play requires reasoning about the global escape structure.
A wall that looks urgent locally may leave several escape corridors open, while a farther wall may close a bottleneck, remove multiple reachable exits, or shrink the cat's reachable region.

\textbf{Case: CircleCat, CEL, Episode 51.}
CircleCat, Episode 51, shows this failure recurring after prior reflection.
The agent entered the episode with \texttt{strategy\_before\_len = 20965}, meaning that it had already updated its strategy after the Episode 50 failure.
The result was still \texttt{score = 0}, and the strategy grew again to \texttt{strategy\_after\_len = 21670}.

In step 7, the cat was located at \texttt{(2,7)} in the upper-right region.
The agent correctly identified that the nearest escape routes were toward the top and right edges.
By step 9, the cat had moved to \texttt{(1,6)}, very close to the top boundary.
At that point, two boundary cells, \texttt{(0,5)} and \texttt{(0,6)}, were open.
The agent selected a move that blocked one of these immediate exits, but it explicitly acknowledged that the other exit remained open.
By step 13, the cat reached \texttt{(0,5)}, which placed it on the boundary and made the position effectively lost.

\textbf{Interpretation.}
The agent does recognize danger, but only after the danger has become locally urgent.
By the time the cat reaches row 1 with two open boundary exits, blocking one cell is insufficient.
The correct strategic behavior would have been to reshape the top-right escape structure earlier, before the cat entered the near-boundary zone.
This is a Level 2 evaluation failure because the agent evaluates the most immediate threat but fails to evaluate the global consequences of earlier wall placements.

\subsection{Level 3: Proceduralization Bottleneck}

\paragraph{Failure Analysis 4: Verbal Strategy Without Proceduralization.}

\textbf{Explanation.}
This failure occurs when the agent stores useful strategic principles in natural language, but these principles do not become executable action-selection procedures.
The agent may write down strong strategies such as preserving mobility, sealing shared gateways, cutting multiple shortest paths, or checking immediate wins and blocks.
However, these remain descriptive rather than operational.
They do not become a reliable decision-time structure of the form: recognize a state pattern, trigger a rule, filter candidate actions, verify the consequence, and select the move.

\textbf{Case: CircleCat, CEL, Episode 50.}
CircleCat, Episode 50, provides a strong example.
Before the episode, the CEL agent's strategy text already contained global containment principles.
It stated that the best walls should build continuous structures, close gaps between wall groups, seal gateways shared by multiple shortest paths, and reduce the cat's live escape tree rather than merely occupy cells near the cat.
The metadata reports \texttt{strategy\_before\_len = 18962} and \texttt{strategy\_after\_len = 20965}, showing that the agent had accumulated substantial strategic text.

However, the actual gameplay did not implement these principles early enough.
In step 7, the cat was at \texttt{(2,5)}, and the agent recognized that the cat was already in the upper-middle approach zone.
The agent also recognized that row 1 was only one move away and that row 0 would then be the boundary.
In step 9, the cat reached \texttt{(1,4)}, and only then did the agent place a wall at \texttt{(0,4)} to remove an immediate top-boundary escape.
By step 11, the cat had reached \texttt{(0,5)}, which was already on the boundary and effectively ended the game.

\textbf{Interpretation.}
This case demonstrates that the agent has learned the language of global containment but has not proceduralized it.
The agent knows that it should seal shared gateways and cut multiple shortest paths, but it does not convert this principle into early wall placement.
The strategy remains a textual principle, not a constraint that forces the agent to act before the cat reaches the near-boundary zone.

\paragraph{Failure Analysis 5: Abstraction Grounding Failure.}

\textbf{Explanation.}
This failure occurs when the agent can name abstract strategic concepts, but those concepts are not grounded in concrete state features or action constraints.
For example, an agent may use terms such as mobility, corner safety, shortest paths, feeder bands, bottlenecks, or reachable exits.
However, these concepts do not consistently trigger the right behavior in the relevant state.
The abstraction is linguistically available, but behaviorally under-grounded.

This failure is distinct from not knowing the concept.
The agent may correctly state that mobility is important, or that a shared gateway should be sealed.
The failure is that the concept does not become a state recognizer or a candidate-action constraint.

\textbf{Case: Othello6, ReasoningBank, Episode 51.}
Othello6, Episode 51, illustrates this failure.
The agent's strategy already included corner-related and mobility-related rules.
It contained instructions to audit every edge or corner-adjacent move, and after losing a corner, to switch from shape-based play to a corner race.
The episode metadata reports \texttt{strategy\_before\_len = 15223}, but the final score was still \texttt{11:25}.

In step 29, the agent explicitly described the position as a low-mobility endgame.
It also stated that the key question was which move would preserve Black's ability to keep playing after White's best reply.
This is the correct abstract diagnosis.
However, in step 31 and step 33, Black had no legal placement and had to pass.
The reflection then noted that the game collapsed when Black ran out of legal moves and White finished the remaining edge squares.

\textbf{Interpretation.}
The agent can use the word mobility correctly, but it does not consistently ground mobility in concrete future legal-move counts.
The concept functions as an after-the-fact explanation rather than as a move-selection constraint.
A grounded use of mobility would require the agent to reject actions that leave Black with no legal moves after White's likely reply, unless an immediate tactical gain justifies that risk.
The trace suggests that this grounding is missing.

\subsection{Level 4: Experience-Integration Bottleneck}

\paragraph{Failure Analysis 6: Long-Horizon Credit Assignment Failure.}

\textbf{Explanation.}
This failure occurs when the final loss is caused by decisions made much earlier, but the agent's reflection focuses mainly on the terminal symptom.
In games such as Othello6, an endgame forced pass is usually not the true causal mistake.
It is the visible consequence of earlier decisions that reduced future mobility, exposed unstable frontier discs, or allowed the opponent to secure stable corners.
If the agent only records that it ran out of moves, without identifying which earlier choices caused the collapse, then the next episode may repeat the same structural error.

\textbf{Case: Othello6, ReasoningBank, Episodes 50 and 51.}
Episode 50 shows the first part of this pattern.
The agent already had a memory item instructing it to audit corner adjacency and check open corners using the exact bracket rule.
Nevertheless, the episode ended with a \texttt{12:24} loss.
The later steps included repeated forced passes.
The reflection also acknowledged that the agent often talked about corner denial, but several move choices still allowed White to take stable corners and convert them into a large endgame advantage.

Episode 51 shows that the issue was not resolved.
The agent again described the endgame as low-mobility and identified preservation of future legal moves as the key criterion.
Yet step 31 and step 33 again resulted in Black having no legal placement, and the final score was \texttt{11:25}.
The repeated pattern suggests that the agent did not successfully trace the terminal mobility collapse back to earlier positional concessions.

\textbf{Interpretation.}
This is a Level 4 failure because the problem lies in integrating experience across time.
The agent's reflection correctly names the symptom, namely forced passes and mobility collapse.
However, it does not reliably identify the upstream decisions that produced the symptom.
As a result, the agent continues to enter endgames where no legal moves remain.
Experience is recorded, but causal credit is not assigned to the earlier choices that need to change.

\paragraph{Failure Analysis 7: Memory Retrieval and Compression Failure.}

\textbf{Explanation.}
This failure concerns the self-evolution mechanism itself.
Longer memory does not necessarily imply better adaptation.
As strategy, memory, and reflection grow longer, the agent still needs to retrieve the relevant lesson, compress it into an actionable form, and apply it to the current state.
If memory remains an unstructured collection of natural-language advice, the most important lessons may be diluted by generic principles, repeated observations, or stale heuristics.

\textbf{Case: CircleCat, CEL, Episodes 50 and 51.}
CircleCat Episodes 50 and 51 show this failure clearly.
In Episode 50, the CEL agent's strategy grew from \texttt{strategy\_before\_len = 18962} to \texttt{strategy\_after\_len = 20965} after a failure in which the cat reached the top boundary.
The agent had already written principles about continuous wall structures, shared gateways, and shortest paths.
Yet the episode still failed because the cat moved from \texttt{(2,5)} to \texttt{(1,4)}, and then to \texttt{(0,5)}.

In Episode 51, the agent entered with the updated strategy text, now \texttt{strategy\_before\_len = 20965}.
Nevertheless, a structurally similar failure recurred.
The cat moved from \texttt{(2,7)} to \texttt{(1,6)}, gained two immediate boundary exits, and escaped through \texttt{(0,5)}.
After this episode, the strategy length increased again to \texttt{strategy\_after\_len = 21670}.
The memory became longer, but the behavior did not reliably change.

\textbf{Interpretation.}
This case suggests that self-evolution may produce more text without producing more usable control.
The agent stores the right strategic ideas, and the memory grows after failure, but the next episode still repeats a similar near-boundary emergency.
The missing step is retrieval and compression: the agent does not convert the prior failure into a compact decision rule such as ``when the cat enters the upper approach zone, prioritize closing shared top exits before row 1 becomes reachable.''
Without that conversion, accumulated memory remains descriptive rather than behaviorally effective.

\subsection{Summary}

The detailed cases support the four-level interpretation in the main paper.
Some failures occur at the level of state simulation, where the agent cannot reliably compute the immediate consequences of an action.
Other failures occur at the level of evaluation, where the agent recognizes that future consequences matter but does not compare candidate actions correctly.
Still others occur at the level of proceduralization, where the agent has useful strategic language but does not convert it into action constraints.
Finally, repeated failures across consecutive episodes indicate an experience-integration bottleneck: memory and reflection accumulate, but they do not reliably produce stable experience-sensitive behavioral change.

These traces suggest that current self-evolving agents do not simply lack strategic descriptions.
Rather, they fail at successive stages of converting experience into behavior: simulating states, evaluating futures, proceduralizing strategies, and integrating prior failures into reusable decision rules.

\begin{table}[t]
\centering
\scriptsize
\setlength{\tabcolsep}{2.8pt}
\renewcommand{\arraystretch}{1.12}
\begin{tabular}{lcccccc}
\hline
 & \multicolumn{2}{c}{Four in a Row}& \multicolumn{2}{c}{Othello6}& \multicolumn{2}{c}{CircleCat}\\
Method & First 10 & Final 10 & First 10 & Final 10 & First 10 & Final 10 \\
\hline
Human median & 0.0\% & 10.0\% & 0.0\% & 0.0\% & 20.0\% & 20.0\% \\
CEL & 0.0\% & 0.0\% & 0.0\% & 0.0\% & 0.0\% & 0.0\% \\
EvoTest & 0.0\% & 0.0\% & 0.0\% & 0.0\% & 0.0\% & 0.0\% \\
EvolveR & 10.0\% & 0.0\% & 0.0\% & 0.0\% & 10.0\% & 0.0\% \\
ReasoningBank & 0.0\% & 0.0\% & 0.0\% & 0.0\% & 0.0\% & 0.0\% \\
\hline
\end{tabular}
\caption{Win/loss outcome diagnostic for humans and language agents.
Agent First 10 and Final 10 are computed over episodes 1--10 and 66--75.
Human values are medians over players after per-player first/final window aggregation.
The table shows that outcome-level win rate is sparse in these hard games: most language agents show no visible improvement.
Thus, win/loss alone provides a weak and insensitive signal for experience-sensitive learning.
}
\label{tab:winlose-outcome}
\end{table}
\section{Metric Implementation Details}
\label{app:metric-details}

This appendix specifies the metric implementation used in our experiments.
All metrics are computed from logged action traces by reconstructing the legal action set at each decision state and comparing the observed action with a game-specific greedy baseline.

\subsection{Cross-Game Greedy-to-Global Metrics}

For each state, we assign evaluator scores to all legal actions and use a state-wise softmax to normalize scores within the action set.
\texttt{GVD} is the normalized value difference between the player's action and the best tied greedy action.
\texttt{GTA} is defined only on trap opportunities, where the best legal action exceeds the greedy baseline by a game-specific threshold; success requires choosing an action with higher evaluator score than the greedy baseline.

\subsection{Four in a Row Metrics}

We use the Pascal Pons solver as the action evaluator.
The greedy baseline is a local line-completion heuristic: for each legal column, we simulate the drop and score the move by the larger of the player's longest extended line and the opponent's longest blocked line.
\texttt{GTA} uses \texttt{TRAP\_THRESHOLD = 1}.
For game-specific metrics, \texttt{EAR} is solver-best agreement, \texttt{OCI} measures opening centrality, and \texttt{CPQ} uses \texttt{CPQ\_THRESHOLD = 2} to identify solver-critical positions.

\subsection{Othello6 Metrics}

The greedy baseline is the legal move that flips the most discs immediately.
Legal actions are evaluated by depth-limited minimax with \texttt{MINIMAX\_DEPTH = 5}.
The leaf evaluator combines positional weights, a mobility term $5(|\mathcal{A}_{\mathrm{human}}|-|\mathcal{A}_{\mathrm{bot}}|)$, and an endgame disc-count term $3(n_{\mathrm{human}}-n_{\mathrm{bot}})$ when fewer than ten empty cells remain; terminal outcomes receive scores of approximately $\pm 10000$.
\texttt{CPQ} and \texttt{GTA} use thresholds of \texttt{5.0} for search-critical and greedy-trap states, respectively.

The mobility-control metric is computed after the player's move:
\begin{equation}
  \mathrm{MCI}
  =
  |\mathcal{A}_{\mathrm{human}}(s_{t+1})|
  -
  |\mathcal{A}_{\mathrm{bot}}(s_{t+1})| ,
  \label{eq:appendix-mci}
\end{equation}
Other quantities such as dangerous-square avoidance, blunder severity, and max-flip rate were used during metric exploration but are not included in the main metric table.

\subsection{CircleCat Metrics}

The greedy baseline is the nearest legal wall placement to the cat, with ties resolved by the highest containment score.
Candidate walls are evaluated by recomputing the cat's reachable escape structure with breadth-first search.
The containment score is a weighted sum of escape-path distance gain,
escape-option reduction, and reachable-area reduction:
$S(a)=3.0\,\mathrm{EPD}(a)+1.5\,\mathrm{EPC}(a)+0.05\,\mathrm{CAR}(a)+B(a)$.
Here $B(a)=20.0$ if the wall eliminates all reachable boundary cells,
and $B(a)=0$ otherwise. We treat a state as a greedy-trap opportunity when the best containment score exceeds the nearest-wall greedy baseline by at least 1.0.

\subsection{Trajectory Aggregation and Filtering}

Human games are sorted chronologically per player and aggregated into non-overlapping five-game groups; group curves report the median across player-level group values.
Human three-stage figures use the first three, middle five, and final three games.
For Othello6 greedy curves, unfinished games with fewer than five human moves are removed.
Agent curves use 75 episodes per game, aggregated into five-episode groups; agent three-stage figures use the first three, middle five, and final three episodes.
For the win/loss table, agent First 10 and Final 10 correspond to episodes 1--10 and 66--75, while human values are computed per player and summarized by the median.

\end{document}